\documentclass[12pt]{article}
\usepackage{pst-plot,epsf}
\newcommand{\beq}{\begin{equation}}
\newcommand{\eeq}{\end{equation}}
\newcommand{\bea}{\begin{eqnarray}}
\newcommand{\eea}{\end{eqnarray}}
\newcommand{\nn}{\nonumber}

\newcommand{\epm}{e^+e^-}

\newcommand{\ra}{\rightarrow}

\newcommand{\ttbar}{t\bar{t}}

\begin{document}
\thispagestyle{empty}
\begin{flushright}
November 2008\\
Revised version\\
January 2009\\
\vspace*{1.5cm}
\end{flushright}
\begin{center}
{\LARGE\bf {\tt carlomat}\\[4mm]
A program for automatic computation of lowest order cross sections}\\
\vspace*{2cm}
Karol Ko\l odziej\footnote{Supported by the Polish Ministry of Scientific 
Research and Information 
Technology as a research grant No. N N519 404034 in years 2008--2010 
and by European Community's Marie-Curie Research Training Network under 
contracts MRTN-CT-2006-035482 (FLAVIAnet) and MRTN-CT-2006-035505 
(HEPTOOLS).}$^,$\footnote{E-mail: karol.kolodziej@us.edu.pl}\\[1cm]
{\small\it
Institute of Physics, University of Silesia\\ 
ul. Uniwersytecka 4, PL-40007 Katowice, Poland}\\
\vspace*{3.5cm}
{\bf Abstract}\\
\end{center}
The current version of {\tt carlomat}, a program for automatic 
computation of the lowest order cross sections of multiparticle reactions, is
described.
The program can be used as the Monte Carlo generator of unweighted
events as well.
\vfill
\newpage
{\large \bf PROGRAM SUMMARY}\\[4mm]
{\it Title of program:} {\tt carlomat}

{\it Version:} 1.0 (November 2008)

{\it Catalogue identifier:}

{\it Program obtainable from:} CPC Program Library or on request by e-mail
from the author

{\it Licensing provisions:} none

{\it Computers:} all

{\it Operating systems:} Linux

{\it Programming language used:} {\tt FORTRAN 90/95}




{\it Distribution format:} gzipped tar archive


{\it Nature of physical problem}\\
Description of two particle scattering reactions with possibly up to 10 
particles in the final state with a complete set of the Feynman diagrams
in the lowest order of the Standard Model.

{\it Method of solution}\\
The matrix element for a user specified process and
phase space parametrizations, which are necessary for the 
multi-channel Monte Carlo integration of the lowest order cross sections and
event generation, are generated automatically. Both the electroweak and 
quantum chromodynamics lowest order contributions are taken into 
account. Particle masses are not neglected in the program. 
Matrix elements are calculated numerically with the helicity amplitude
method. Constant widths of unstable particles are implemented by
modifying mass parameters in corresponding propagators. 

{\it Restrictions on complexity of the problem}\\
The number of external particles is limited to 12.
Only the Standard Model is implemented at the moment in the program.
No higher order effects are taken into account, except for assuming 
the fine structure
constant and the strong coupling at appropriate scale and partial summation
of the one particle irreducible loop corrections by introducing fixed widths of
unstable particles.

{\it Typical running time}\\
Generation of the Fortran code with {\tt carlomat} on 
a PC with the Pentium~4 3.0~GHz processor for reactions with 8 particles
in the final state relevant for the associated associated top quark pair 
and Higgs boson production and decay takes about 10 minutes CPU time. 
This relatively long time 
of the code generation is determined by a lot of write to and read from a disk
commands which have to be introduced in order to circumvent limitations of 
the Fortran compilers concerning possible array sizes. The compilation 
time of generated routines depends
strongly on a compiler used and an optimization option chosen. Typically,
for the reactions mentioned, it takes about one hour to compile all the
routines generated. Most of the time is used for 
the compilation of the kinematical routines.
The execution time of the Monte Carlo (MC) integration with about 
2 million calls 
to the integrand amounts typically to a few hours and, if the MC summing 
over polarizations is
employed, it is dominated by computation of the phase space normalization.

\vspace*{1.5cm}
{\large \bf LONG WRITE-UP}\\
\section{Introduction}
Many interesting aspects of the Standard Model (SM) and models beyond it
can be studied through investigation of reactions involving a few
heavy particles at a time whose observation becomes possible 
owing to the increasing collision energy and luminosity of particle colliders,
such as, e.g., the Large Hadron Collider (LHC), or the 
International Linear Collider (ILC) \cite{ILC}.
As the heavy particles are usually unstable,
they almost immediately decay leading to reactions with several
light particles in the final state which receive 
contributions typically from many thousands of Feynman diagrams
already in the lowest order of SM. Although, in their overwhelming majority,
those diagrams constitute background to the ``signal diagrams'' 
of production and decay of the heavy particles one wants to investigate, 
they must be all taken into account in reliable SM 
predictions for such reactions. 
This can be done practically only through a fully automated 
calculational process. 

\begin{figure*}[htb]
\vspace{4cm}
\includegraphics{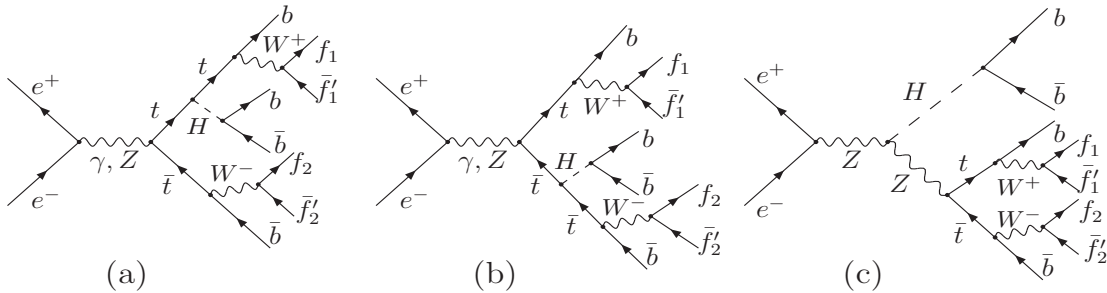}
\caption{Representative signal Feynman diagrams of reaction (\ref{ee8f}) in 
the unitary gauge.
The remaining diagrams are obtained by all possible permutations of 
the two $b$ and two $\bar b$ lines. The Higgs boson coupling to electrons
has been neglected.}
\label{fig:ee8f}
\end{figure*}

To be more specific, let us consider, e.g. a reaction of associated 
production of a top quark pair and a Higgs boson at the ILC 
\beq
\label{eetth}\epm \ra t \bar t H
\eeq
that can be used to measure the top--Higgs Yukawa coupling \cite{eetth}.
The top and antitop decay, even before they hadronize, predominantly into 
$b W^+$ and $\bar{b} W^-$, respectively, 
the electroweak (EW)  bosons subsequently decay into a 
fermion--antifermion pair each
and the Higgs boson, if it has mass $m_H < 140$~GeV, which is favoured by 
the direct searches at LEP 
and theoretical constrains in the framework of the SM,
decays mostly into a $b \bar b$-quark 
pair. Thus, reaction (\ref{eetth}) will actually be detected through 
reactions of the form
\beq
\label{ee8f}
  e^+e^-\;\; \ra \;\;  b \bar b  b \bar b f_1\bar{f'_1} f_2 \bar{f'_2},
\eeq
where $f_1, f'_2 =\nu_{e}, \nu_{\mu}, \nu_{\tau}, u, c$ and 
$f'_1, f_2 = e^-, \mu^-, \tau^-, d, s$ are the decay products  of the 
$W$-bosons coming from decays of the $t$- and $\bar t$-quark. 
Except for 20 `signal' Feynman diagrams of the associated production of a top 
quark pair and a Higgs boson the representatives of which are depicted
in Fig.~\ref{fig:ee8f}, there are many other, typically 
$\mathcal{O}\left(10^4\right)$, off resonance background diagrams which 
contribute to any of reactions (\ref{ee8f}).
For example, taking into account both the EW and quantum 
chromodynamics (QCD) lowest order contributions in the unitary
gauge, neglecting of the Yukawa couplings
of the fermions lighter than $c$ quark and $\tau$ lepton, 
reactions
\bea
\label{tnmn}
\epm &\ra& b \bar b b \bar b \tau^+ \nu_{\tau} \mu^- \bar \nu_{\mu},\\
\label{udmn}
\epm &\ra& b \bar b b \bar b u\bar d \mu^- \bar \nu_{\mu},\\
\label{udsc}
\epm &\ra& b \bar b b \bar b u\bar s d \bar c, \\
\label{uddu}
\epm &\ra& b \bar b b \bar b u\bar d d \bar u
\eea
receive contributions from 21\,214, 26\,816 and 39\,342 and 185\,074
Feynman diagrams, respectively. To which extent the background contributions
may affect the associated production of the top quark pair and
Higgs boson in some of the reactions listed above has been discussed in
\cite{KS}.

Although there exists several multipurpose Monte Carlo (MC)
generators such as {\tt HELAC/ PHEGAS}~\cite{HELAC/PHEGAS}, 
{\tt AMEGIC++/Sherpa}~\cite{AMEGIC++/Sherpa}, 
{\tt O'Mega/Whizard}~\cite{O'Mega/Whizard},
{\tt MadGraph/MadEvent}~\cite{MadGraph/MadEvent}, {\tt ALPGEN}~\cite{ALPGEN}, 
{\tt CompHEP/CalcHEP}~\cite{CompHEP}, or recently released 
{\tt Comix}~\cite{Comix}, 
three years ago, when work on {\tt carlomat} started, none of
them was able to handle multiparticle reactions 
such as (\ref{tnmn})--(\ref{udsc}), not to mention reaction (\ref{uddu}).
This situation has changed quite recently and now some 
of the publicly available MC generators, 
as e.g. {\tt O'Mega/Whizard} or {\tt HELAC/ PHEGAS}, are able to deliver 
complete SM predictions for reactions with 8 and more particles in the final 
state, including both the EW and QCD contributions 
in the leading order. However, when first physics results 
obtained with {\tt carlomat}, on the off resonance
background in (\ref{eetth}) \cite{KS},
were submitted for publication in spring 2008, they could only be partially 
checked against generators that were publicly available then.

The main aim of {\tt carlomat}, a new multipurpose program for 
automatic computation of the lowest order cross sections, is to provide
the reliable description of multiparticle reactions. This is a very 
challenging task, not only because of complicated matrix elements,
involving many thousands of the Feynman diagrams, but also because of
the necessity of performing integration over a multidimensional phase
space. Not only may {\tt carlomat} be useful for tests of the complicated
calculations done with other, already well established, generators.
It may also offer the user other advantages, such as
a relatively high speed of the MC computation, an easy control over
mappings of peaks in the integrand, 
a possibility of calculating
polarized cross sections, or a possibility of easy hand made modifications 
due to the use of traditional Feynman diagram approach.
This write-up describes the current version of the program.

\section{Basics features of the program}

{\tt carlomat} is a program written in Fortran 90/95.
It generates the matrix element for a user specified process and
phase space parametrizations which are later used for the 
multichannel MC integration of the lowest order cross sections and
event generation. The program takes into account both the EW and QCD 
lowest order contributions. Particle masses are not neglected in
the program. The number of external particles is limited to 12 and
only the SM is currently implemented in the program.

{\tt carlomat} works according to the following scheme.
The user specifies the process he/she wants to have calculated.
Then topologies for a given number of external particles are generated
and checked against Feynman rules which have been coded in the program. 
In this process,
helicity amplitudes, the colour matrix and phase space parametrizations are
generated.
Finally, they are copied to another directory where the numerical program 
can be executed.

\subsection{Generation of topologies}

\begin{figure*}[htb]
\vspace{3.cm}
\includegraphics{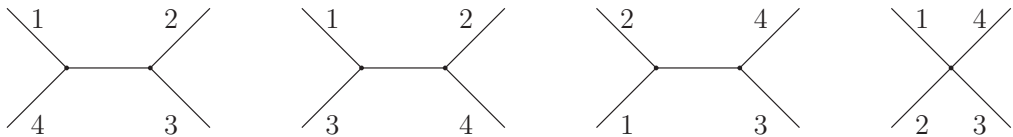}
\vspace*{-1.0cm}
\caption{The topologies of a 4 particle process.}
\label{fig:top4}
\end{figure*}

Let us consider models with triple and quartic couplings.
Generation of topologies starts 
with 4 topologies of a 4 particle process which are depicted in 
Fig.~\ref{fig:top4}. The 25 topologies of a 5 particle process are obtained 
by attaching line No. 5 to each line, 
including the internal ones, and to each triple vertex of the graphs 
in Fig.~\ref{fig:top4}.
Particles No. 6, 7 and so on, are adeed recursively in the same way.
The number of topologies grows dramatically with the number of external
particles as can be seen from Table~\ref{table:tops}.
\begin{table}[htb]
\begin{center}
\caption{No. of topologies for a given No. of particles.}
\vspace*{3mm}
\label{table:tops}
\begin{tabular}{cr}
\hline\\
No. of particles & No. of topologies\\[2mm]
\hline\\
6  & 220 \\
7  & 2\,485 \\
8  & 34\,300\\
9  & 559\,405\\
10\,\,\,\, & 10\,525\,900\\
11\,\,\,\, & 224\,449\,225
\end{tabular}
\end{center}
\end{table}
However, for a process with $n$ external particles, it is enough to
generate topologies for $n - 1$ particles 
and then, while adding the $n$-th particle, 
to check whether a topology results in a Feynman diagram or not.
Topologies can be generated and stored on a disk prior to 
the program execution.

\subsection{Feynman diagrams}
Actual initial and final state particles are assigned to lines $1,2,3,...,n$ 
in a strict order. Each topology is divided into two parts which are 
separately checked against the Feynman rules. While doing so,
two, or three, external lines are joined by means of a triple, or
quartic, vertex of the implemented model, respectively.
In this way a new off shell particle is formed. Another particle is formed
by adding either the next external line, or an off shell particle that has been
already formed. At the stage of matrix element generation, such 
particles are represented in {\tt carlomat} by the derived data type called 
{\tt particle} that is defined in the following way

      {\tt
      type particle\\
      integer :: wr,ct\\
      integer :: k(4)\\
      integer, pointer :: ip(:)\\
      end type particle,\\[3mm]}
where\\ 
{\tt wr=1/0} if the particle has already/not yet been written to a file,\\
{\tt ct=1/3/8} for a colour singlet/triplet/octet state,\\
{\tt k(1)} -- the number of the vertex that has been used to form 
the particle,\\
{\tt k(2),k(3),k(4)} -- the numbers of particles that have been joined
in the vertex, with {\tt k(4)=0} for a triple vertex,\\
{\tt ip(:)} --  a list of all the off-shell particles that have been used in
the definition of the current particle.\\[2mm]
In the routine for computation of the matrix element, the particles 
being fermions, vector bosons, or scalar bosons are 
represented by spinors, polarization vectors, or just complex numbers, 
respectively.

The off shell particles and/or external particles are joined in this way
until the two parts of a considered topology are completely covered.
If they match into a propagator of the implemented model then the topology 
is accepted.
Once the topology has been accepted, the `longer' part of it is further 
divided so that the Feynman diagram is made of 3 or 4 parts, joint to form
a triple or quartic vertex of the model. This reduces the number of
different particles that must be defined for all the Feynman diagrams.

When the diagram is created, the corresponding particles
are used to construct the helicity amplitude, 
colour factor (matrix) and phase space parametrization which
are stored on the disk.
Once all the topologies have been checked 
subroutines for calculating the matrix element, colour matrix and phase space
integration are written.

\subsection{Helicity amplitudes}

Routines for computing the helicity amplitudes have been collected in
a separate directory named {\tt carlolib}. Most of the routines in
{\tt carlolib} had been already written for an MC program 
{\tt eett6f} \cite{eett6f} for calculating lowest order cross sections of 
reactions $\epm \ra $ 6 fermions, relevant for $\ttbar$-pair production 
and decay, 
but they have been improved and tailored to 
meet the needs of the automatic generation of amplitudes in {\tt carlomat}.
In order to speed up the computation the MC summing over helicities has
been implemented in the program. To illustrate how it works
let us consider a sum over two polarization states $\lambda=\pm 1$
\bea
\label{polsum}
\sum_{\lambda=\pm 1}M^{*}\left(\lambda\right)M\left(\lambda\right)
=\sum_{\lambda=\pm 1}u_a^{\dag}\left(\lambda\right)
\mathcal{O}_a^{\dag}\mathcal{O}_bu_b\left(\lambda\right),
\eea
where the polarized amplitude that depends on $\lambda$ has been written
as
\bea
\label{polamp}
M\left(\lambda\right)=\mathcal{O}_au_a\left(\lambda\right).
\eea
In Eq.~(\ref{polamp}), $u\left(\lambda\right)\equiv 
u\left({\rm\bf p},\lambda\right)$ is the spinor representing a particle
or an antiparticle of momentum {\bf p} and helicity $\lambda/2$,  $\mathcal{O}$
stands for the rest of amplitude $M\left(\lambda\right)$ that does not depend
on $\lambda$ and a summation over the Dirac index $a=1,2,3,4$ has been assumed.
Now let us define the spinor
\bea
\label{spinor}
w\left(\alpha\right)=e^{i\alpha}u\left(+1\right)+e^{-i\alpha}u\left(-1\right),
\eea
with $\alpha$ being a random number uniformly distributed in the interval
$[0,1]$. Then the sum over polarizations can be replaced with 
the integral
\bea
\label{intpol}
\sum_{\lambda=\pm 1}M^{\dag}\left(\lambda\right)M\left(\lambda\right)=
\frac{1}{2\pi}\int\limits_0^{2\pi} w^{\dag}_a\left(\alpha\right)
\mathcal{O}_a^{\dag}\mathcal{O}_bw_b\left(\alpha\right)
\;{\rm d}\alpha.
\eea
Indeed, 
\bea
& &\frac{1}{2\pi}\int\limits_0^{2\pi} w^{\dag}_a\left(\alpha\right)
\mathcal{O}_a^{\dag}\mathcal{O}_bw_b\left(\alpha\right) \;{\rm d}\alpha = \nn\\
& &\frac{1}{2\pi}\int\limits_0^{2\pi} \left(
e^{-i\alpha}u^{\dag}_a\left(+1\right)+e^{i\alpha}u^{\dag}_a\left(-1\right)\right)
\mathcal{O}_a^{\dag}\mathcal{O}_b\left(e^{i\alpha}u_b\left(+1\right)
+e^{-i\alpha}u^{\dag}_b\left(-1\right)\right){\rm d}\alpha=\nn\\
& &u^{\dag}_a\left(+1\right)\mathcal{O}_a^{\dag}\mathcal{O}_bu_b\left(+1\right)
+u^{\dag}_a\left(-1\right)\mathcal{O}_a^{\dag}\mathcal{O}_bu_b\left(-1\right)=
\sum_{\lambda=\pm 1}M^{\dag}\left(\lambda\right)M\left(\lambda\right),\nn
\eea
as $\int\limits_0^{2\pi} e^{2i\alpha}{\rm d}\alpha =
\int\limits_0^{2\pi} \; e^{-2i\alpha}{\rm d}\alpha=0$ and 
$\int\limits_0^{2\pi}{\rm d}\alpha=2\pi$.

Analogously, a sum over three polarization states 
can be replaced with the integral according to
\bea
\label{intpol1}
\sum_{\lambda=0,\pm 1}M^{*}\left(\lambda\right)M\left(\lambda\right)=
\frac{1}{2\pi}\int\limits_0^{2\pi}\epsilon^{*}_{\mu}\left(\alpha\right)
\mathcal{O}^{\mu\;\dag}\mathcal{O}^{\nu}\epsilon_{\nu}\left(\alpha\right)
{\rm d}\alpha,
\eea
where the longitudinal polarization
component must be taken into account in the definition of 
$\epsilon\left(\alpha\right)$
\bea
\label{polvect}
\epsilon\left(\alpha\right)=\varepsilon\left(0\right)+
e^{i\alpha}\varepsilon\left(+1\right)
+e^{-i\alpha}\varepsilon\left(-1\right)
\eea
in terms of the polarization four vectors $\varepsilon\left(\lambda\right)\equiv
\varepsilon\left({\rm\bf p},\lambda\right)$, $\lambda=0,\pm 1$, 
of a massive vector boson.
Eq.~(\ref{intpol1}) can be proven in exactly the same way as 
Eq.~(\ref{intpol}).

Although random sampling over external helicities increases the dimension of 
the integral, by one for each helicity,
it does not actually spoil its convergence,
as the dependence on $\alpha$ in (\ref{intpol}) or (\ref{intpol1})
is rather smooth and the efficiency of the MC integration does not depend
very much on the dimension of the integral. This has been explicitly
checked for a number of reactions $\epm\ra 4\;{\rm fermions}\;(\gamma)$
whose cross sections calculated 
with a program {\tt ee4f$\gamma$} \cite{ee4fgamma} that applies
explicit summing over helicities have been successfully reproduced 
with {\tt carlomat} using the random sampling over external helicities.

An explicit summing over helicities is also possible. While doing so,
spinors or polarization vectors representing particles, both the on- and
off-shell ones, are computed only once, for all the helicities of the external
particles they are made of, and stored in arrays, which are later used
in the sum over polarizations.

Possible poles in the propagators of unstable particles are regularized 
by their constant widths which 
are introduced through the complex mass parameters
\bea
\label{cmass}
M_B^2\!&=\!&m_B^2-im_B\Gamma_B, \qquad B=W, Z, H, \\
 \qquad M_t&=&\sqrt{m_t^2-im_t\Gamma_t},\nn
\eea
which replace masses in the corresponding propagators
\bea
\Delta_F^{\mu\nu}(q)\!&=&\!\frac{-g^{\mu\nu}+q^{\mu}q^{\nu}/M_V^2}
                               {q^2-M_V^2},   \nonumber \\
\Delta_F(q)\!&=&\!\frac{1}{q^2-M_H^2}, \qquad
S_F(q)=\frac{/\!\!\!q+M_t}{q^2-M_t^2}, \nonumber
\eea
both in the $s$- and $t$-channels.
Propagators of a photon and gluon are taken in the Feynman gauge.
The EW mixing parameter may be defined either to be real
\bea
\label{FWS}
\sin^2\theta_W=1-\frac{m_W^2}{m_Z^2},
\eea
with physical values of the $W$ and $Z$ boson masses, or complex
\bea
\label{CMS}
\sin^2\theta_W=1-\frac{M_W^2}{M_Z^2}, 
\eea
with $M_W^2$ and $M_Z^2$ given by Eq.~(\ref{cmass}). Using definitions
(\ref{FWS}), or (\ref{CMS}), in the EW coupling constants is usually
referred to as the fixed width scheme (FWS), or the complex 
mass scheme (CMS) \cite{Racoon}.
The colour matrix is calculated only once at the beginning of execution 
of the numerical program after having reduced its size with the use of 
the SU(3) algebra properties. 
The calculation is performed numerically from the very definition, 
using the SU(3)  
structure constants and group generators in the fundamental representation
\cite{Pokorski}.

\subsection{Phase space integration}

The integration over a multidimensional phase space is practically possible
only with the use of Monte Carlo methods. A number of approaches has been 
developed in the literature to reduce the variance of the integrand
\cite{varred}. The multichannel MC approach that is
used in {\tt carlomat} is described in this section.

A dedicated phase space parametrization is generated for each Feynman diagram
of a process
\bea
\label{momcons}
p_1 +p_2 \ra p_3 + ... + p_n
\eea
with $n$ external particles of four momenta $p_i$, $i=1,2,...,n$,
in the centre of mass system.
The Lorentz invariant phase space element of (\ref{momcons})
is defined in a standard way
\bea
\label{dpsgen}
 {\rm d}^{3n_f-4} Lips  &=& (2\pi)^{4} \delta^{(4)}\left(p_1 +p_2 
- \sum_{i=3}^n p_i\right) \prod_{i=3}^n\frac{{\rm d} p_i^3}{(2\pi)^32E_i},
\eea
where $n_f=n-2$ is the number of particles in the final state.
The set of final state particles is divided, in a way that depends 
on a topology of the diagram, into subsets consisting of
two elements each, an element being either an external particle or
a subset of particles itself and use is made of the identity
\bea
\int{\rm d}s_i\int\frac{{\rm d}^3 q_i}{2E_i} \; 
\delta^{(4)}\left(q_i-q_{i_1}-q_{i_2}\right)=1, 
\qquad E_i^2=s_i+{\rm \bf q}^2_i.\nn
\eea
Thus, Eq.~(\ref{dpsgen}) can be brought into the following form
\bea
\label{dps}
 {\rm d}^{3n_f-4} Lips  &=& (2\pi)^{4-3n_f}  
          {\rm d} l_{0} {\rm d} l_{1}...{\rm d} l_{n-4}
          {\rm d} s_{1} {\rm d} s_{2}...{\rm d} s_{n-4},
\eea
where ${\rm d} l_i$, $i=0,1,...,n-4$, is a two particle phase space 
element given by
\bea
\label{dps2pt}
{\rm d} l_i=\frac{\lambda^{\frac{1}{2}}
\left(s_i,q_{i_1}^2,q_{i_2}^2\right)}{2\sqrt{s_i}}
 {\rm d} \Omega_i.
\eea
In Eq.~(\ref{dps2pt}), $\lambda$ is the kinematical function, 
$q_{i_1}$ and $q_{i_2}$ are the four momenta of each 
particle (subsystem of particles) in subset $i$, which are defined in 
their relative centre of mass system, 
${\bf q}_{i_1}+{\bf q}_{i_2}={\bf 0}$,  $\Omega_i$ is the solid angle
of momentum ${\bf q}_{i_1}$ and invariants $s_i$ are given by
\bea
\label{invs}
s_i=\left\{ \begin{array}{l}
\left(q_{i_1}+q_{i_2}\right)^2=\left(E_{i_1}+E_{i_2}\right)^2, \quad {\rm for}
\;\; i=1,...,n-4 \\
\left(p_1+p_2\right)^2=s, \quad {\rm for}\;\; i=0 
\end{array} \right..
\eea
For processes with identical particles in the initial and final states, 
parametrization (\ref{dps}) must be slightly modified, by introducing 
the corresponding $t$ invariants in a proper way. This has not 
been implemented in the program yet. Therefore, one may expect worse
convergence of the MC integration when {\tt carlomat} is used 
for such processes. The convergence
is not much worse, however, if an angular cut
of a few degree on the angle between  the identical particles in the 
initial and final states is imposed, which has been tested for several
reactions $\epm\ra 4\;{\rm fermions}$ and $\epm\ra 6\;{\rm fermions}$
with electrons and/or positrons in the final state.

Invariants $s_i$ of Eq.~(\ref{invs}) are randomly generated within their
physical
limits, $s_i^{\rm min}$ and $s_i^{\rm max}$, which are automatically deduced 
from a topology of the Feynman diagram. They can be generated either
according to the uniform distribution
\beq
\label{si}
       s_i=\left(s_i^{\rm max} - s_i^{\rm min}\right) x_i + s_i^{\rm min},
\eeq
where $x_i$ denotes a random variable uniformly
distributed in the interval $\left[0,1\right]$,
or, if necessary, mappings of
the Breit-Wigner shape of the propagators of unstable particles and
$\sim 1/s$ behaviour of the propagators of massless particles
are performed. An option is included in the program that allows to
turn on the mapping if the particle decays into 2, 3, 4, ...
on-shell particles. Different phase space parametrizations obtained in this
way are important for testing purposes.


Phase space parametrizations (\ref{dps}) generated for each Feynman diagram 
\bea
f_i(x)={\rm d}^{3n_f-4} Lips_i\left(x\right) \qquad i=1,...,N,\nn
\eea
with $N$ being the number of the diagrams and 
$x=\left(x_1,...,x_{3n_f-4}\right)$
uniform random arguments, satisfy the normalization condition
\bea
\label{norm}
\quad \int\limits_0^{\; 1}{\rm d} x^{3n_f-4}f_i(x)=1.
\eea
They are used to define a new multichannel probability distribution 
\bea
\label{distrib}
f(x)=\sum_{i=1}^N a_i f_i(x), 
\eea
with non negative weights $a_i$, $i=1,...,N$, satisfying the condition
\bea
\label{norma}
\sum_{i=1}^N a_i = 1
\eea
which guaranties that the combined distribution $f(x)$ is normalized
if every distribution $f_i(x)$ satisfies normalization condition (\ref{norm}).
The actual MC integration is done with the random numbers generated
according to probability distribution $f(x)$. 
An option has been included in the program that allows for reducing
the actual number of kinematical channels used in the integration.
It can be turned on by choosing {\tt iopkch=1}
in {\tt carlocom.f}. Each kinematical channel is then called with the same
set of random arguments $x$ and of all the weights of the channels with 
the same phase space  normalization $f_i(x)$ only one weight is kept nonzero.

The integration is performed iteratively. The first iteration 
starts with all the nonzero weights $a_i$ equal to each other,
or with the weights determined in the initial scan of the integrand
according to the following formula
\beq
\label{ai}
          a_i=\sigma_i/\sum_{j=1}^N \sigma_j,
\eeq
where $\sigma_j$ denotes the cross section obtained with the $i$-th 
kinematical channel. An option has been included in the program that
allows to choose whether the weights should be determined anew according 
to Eq.~(\ref{ai}) after every iteration, or a fraction of old weights, i.e.
determined in the scan or previous iterations, should be preserved and
transfered to the next iteration.
This is controlled by specifying the fraction {\tt po} in {\tt carlocom.f},
as described in Section~3.2. 


%
\section{Description of the program}
The two basic parts of the program
which allow for a generation of the Fortran code and 
the MC computation of the cross section of a user defined process
are described in the following two subsections.
The default values of the input parameters and options used in the program
are those specified below.

\subsection{Fortran code generation}

The part of the program that is responsible for generating 
routines necessary for the execution of the MC program is 
stored in a directory {\tt code\_generation}. 
The main program is stored in a file {\tt carlomat.f}.
The user should specify the process
he wants to have calculated by giving a value to a character variable
{\tt process} in  {\tt carlomat.f}, e.g.\\[2mm]
{\tt      process='e+ e- -> b b\~\,mu+ vm d u\~\,'}.\\[2mm]
He may also choose the following options in {\tt carlomat.f}:\\[2mm]
Should the MC summing over polarizations be done, 
{\tt imcs=1 (yes)/else (no)}?\\[2mm]
{\tt      imcs=1.} {\em Recommended for multiparticle reactions.}\\[2mm]
Take into account pure EW contributions only, 
      {\tt iewk=1(yes)/else(no)}?\\[2mm]
{\tt      iewk=0.} {\em Recommended; takes into account both QCD and EW 
                   contributions.}
\\[2mm]
Discard the Higgs boson contributions, {\tt ihgs=1(yes)/else(no)}?\\[2mm]
{\tt      ihgs=0.} {\em Recommended; Higgs boson contributions are taken 
                    into account.}

Then the program calls two subroutines: {\tt readpart} and {\tt genmat}.

In {\tt readpart}, the initial and final
state particles are read from {\tt process}. They are compared with the 
particles of the implemented model stored in a file {\tt modpart.dat} and
assigned the necessary characteristics which will be used during execution
of the program. If all the particles match with particles in 
{\tt modpart.dat} the possible interactions are read from a file 
{\tt vertices.dat}. The numbers of vertices of different kind should
not exceed the corresponding maximum numbers specified in a module 
file {\tt maxvals.f}.
Except for the maximum numbers of vertices:\\[2mm]
{\tt nffb=48,} the number of fermion--fermion--boson vertices, \\[2mm]
{\tt nbbb=6,} the number of triple boson vertices,\\[2mm]
{\tt nbbbb=8,} the number of quartic boson vertices\\[2mm]
{\tt maxvals.f} contains also values of other parameters which 
allow to control execution of the program:\\[2mm]
{\tt ncol=3,} the number of colors,\\[2mm]
{\tt mxprt=100,} the maximum number of particles in a model,\\[2mm]
{\tt mxepl=12,} the maximum number of external particles,\\[2mm]
{\tt mxipl=100000,} the maximum number of the off-shell particles
                   which can be created,\\[2mm]
{\tt mxvpl=4,} the maximum number of similar vertices, such as $uu\gamma$, 
$uuZ$, $uuh$, or $uug$,\\[2mm]
{\tt mxncnt=20,} the maximum number of continuation lines in a command,\\[2mm]
{\tt mxcmdl=1000,} the maximum number of lines in selected routines as, 
e.g., {\tt mdkk*.f}, {\tt mmodkk*.f},\\[2mm]
{\tt mxcmdb=20,} the maximum number of blocks in selected routines as, e.g.,
{\tt kinkk*.f}, relevant for the speed of their compilation,\\[2mm]
{\tt mxmap=4,} the maximum number of particles in a subsystem whose 
propagator should be mapped.\\[2mm]
The module contains also the name of the current version of the program.

In {\tt genmat}, topologies for the actual number of $n_{\rm ext}$ external 
particles are generated, if the corresponding file 
containing pregenerated topologies for $n=n_{\rm ext}-1$ particles
could not have been found in a  directory {\tt code\_generation}.
This is done by consecutive calls to a subroutine {\tt gentop} with
the number of particles $n=4,5,...,n_{\rm ext}$. As long as $n < n_{\rm ext}$
the topologies are being generated and stored in a file named {\tt topoln.dat}. 
When {\tt gentop} is called for the last time, with $n = n_{\rm ext}$, 
each time a new topology is created by attaching the last particle 
to a topology that was generated in the previous call and that has been 
now read from a file {\tt topoln-1.dat}, 
another subroutine {\tt checktop} is called,
which checks it against Feynman rules in a way described in Section~2.2.
While doing so, the off-shell particles are created which,
if the topology has been accepted,  are used to generate the
helicity amplitudes, the colour matrix and phase space parametrizations
for all the Feynman diagrams corresponding to the topology.

After all the topologies have been inspected, the following files necessary
for a computation of the matrix element and phase space integration are 
created and stored in the working directory. 
\begin{itemize}
\item Modules:\\[2mm]
{\tt partcls.f}, containing information about the process, as, e.g.,
                 numbers of particles, diagrams, etc.,\\[2mm]
{\tt umodkk.f}, a collection of complex arrays representing spinors,
                polarization vectors and scalars which are the building 
                blocks of the diagrams,\\[2mm]
{\tt hmodkk.f}, a collection of integer arrays which determine a sequence
                of elements in arrays of {\tt umodkk}, relevant
                if no MC summing over polarizations is done,\\[2mm]
{\tt pmodkk.f}, a collection of particle four momenta,\\[2mm]
{\tt mmodkk*.f}, a collection of the amplitudes.
\item Subroutines:\\[2mm]
{\tt matrixel.f}, the squared matrix element averaged (summed) over 
                  polarizations and colors of the initial (final) state
                  particles,\\[2mm]
{\tt mpol2.f}, the polarized squared matrix element averaged (summed) over 
               colors of the initial (final) state particles,\\[2mm]
{\tt mtel*.f}, a collection of calls to routines of {\tt carlolib} which
               calculate building blocks of the Feynman diagrams,\\[2mm]
{\tt mdkk*.f}, a collection of calls to routines of {\tt carlolib} which
               calculate amplitudes of the Feynman diagrams,\\[2mm]
{\tt hsubkk.f}, calculates integer arrays which determine a sequence
                of elements in arrays of {\tt umodkk}, relevant
                if no MC summing over polarizations is done,\\[2mm]
{\tt colsqkk.f}, calculates colour factors,\\[2mm]
{\tt kinkk*.f}, a collection of {\tt mxcmdb} different phase space
                parametrizations,\\[2mm]
{\tt  kincls.f}, calls the kinematical routines {\tt kinkk*} in order
                 to generate the final state particle four momenta,\\[2mm]
{\tt psnkk.f}, calls phase space normalization routines {\tt psub*},\\[2mm] 
{\tt psub*.f}, calls the kinematical routines {\tt kinkk*} in order
               to calculate the phase space normalization.
\item Parts of the code:\\[2mm]
{\tt partcls.dat}, some information about the external particle of the process,
                   included in a subroutine {\tt parfixkk},\\[2mm]
{\tt mtdr*.f}, adds up the amplitudes which have the same colour factors,
               included in a subroutine {\tt mpol2}.
\end{itemize}
Once the execution of {\tt carlomat} ends 
they are copied to another directory named {\tt mc\_computation}, where 
the MC program can be executed.

\subsection{MC computation}

The part of the program that is responsible for the MC computation
is stored in a directory {\tt mc\_computation}. 
The main program is stored in a file {\tt carlocom.f}.
In the latter, the user should supply the following data:\\[2mm]
The centre of mass energies\\[2mm]
{\tt aecm=(/500.d0/)}, \\[2mm]
for which the cross section should be computed together with a parameter\\[2mm]
{\tt ne=1}\\[2mm]
being a size of the array {\tt aecm}. If, for example, the program should
run for 3 centre of mass energies: 500 GeV, 800 GeV and 1 TeV then
{\tt aecm=(/500.d0,800.d0,1.d3/)} and {\tt ne=3}.
Moreover, the following
options should be chosen in {\tt carlocom.f}.\\[2mm]
Should the Born cross section be calculated, {\tt iborn=1(yes)/else(no),}
in {\tt niter} iterations with {\tt ncalli} calls per iteration, i.e. 
with {\tt niter} $\times$ {\tt ncalli} calls?\\[2mm]
{\tt      iborn=1,} {\em recommended,}\\
{\tt      niter=10,}\\
{\tt      ncalli=10000.}\\[2mm]
The switch {\tt iborn} has been introduced, since the implementation 
of the initial state radiation within the structure function approach 
is envisaged. Its only meaningful value in the current version of the
program is 1.\\
The fraction of old integration weights which should
be transfered to the next iteration\\[2mm]
{\tt      po=0.d0,}  {\em recommended value from 0 to 0.5.}\\[2mm]
If {\tt po=0.d0} then integrations weights $a_i$ are
calculated anew according to Eq.~(\ref{ai}) and it may happen that, because
of a low weight determined in the previous iteration, the $i$-th channel
will not be selected resulting in $a_i=0$ for the next iteration. 
Thus the number of kinematical
channels used in the next iterations will be effectively reduced.\\[2mm]
Generate the unweighted events, {\tt imc<0(no)/else(yes)}?\\[2mm]
{\tt      imc=2.}\\[2mm]
The probability distribution according to which the momenta of the final 
state particles
are generated is limited either by the maximum of the total cross section,
for {\tt imc=0}, or, for {\tt imc=1,2,...}, by the maxima of the 
cross section in each bin of distribution No. {\tt imc}. See discussion
of an option {\tt idis} below.\\[2mm]
Should the Born cross section be scanned with {\tt nscan0} calls,
{\tt iscan0=1(yes)/else(no)}?\\[2mm]
{\tt      iscan0=1,}\\
{\tt      nscan0=10.}\\[2mm]
If {\tt iscan0=1} then each kinematical channel is called with the same
weight equal to $1/n_{\rm kin}^{\rm tot}$, where $n_{\rm kin}^{\rm tot}$
is the total number of the channels, and the cross section is 
calculated with {\tt nscan0} calls. The value of the cross section
is later used to calculate the weight, according to Eq.~(\ref{ai}), with which 
that particular channel will contribute to the integral in the first iteration.

Should the number of kinematical channels be optimized, 
{\tt iopkch=1(yes)/else(no)}?\\[2mm]
{\tt  iopkch=0,} {\em for processes with 
large number of the Feynman diagrams }{\tt  iopkch=1} 
{\em is recommended.}\\[2mm]
Discard kinematical channels which contribute less than {\tt qw} of
the dominant channel in the initial scan.\\[2mm]
{\tt  qw=0.0d0.} {\em No channels are discarded; recommended.}\\[2mm]
Choose the scheme, {\tt ischeme=1(complex mass scheme)/else(fixed width 
scheme)}, see Eqs.~(\ref{FWS}) and (\ref{CMS}) for the explanation,\\[2mm]
{\tt  ischeme=1.}\\[2mm]
Should distributions be calculated, {\tt idis=1(yes)/0(no)}?\\[2mm]
{\tt      idis=1.}\\[2mm]
Variables, in which the distributions are to be calculated, the number
of bins, etc., should be defined in a file {\tt calcdis.f}.\\[2mm]
Should cuts be imposed, {\tt icuts=1(yes)/0(no)}?\\[2mm]
{\tt    icuts=1.}\\[2mm]
The cuts which can be controlled from the main program {\tt carlocom} 
currently include
\begin{itemize}
\item angles:
$\theta\left({\rm l},{\rm beam}\right)$, 
$\theta\left({\rm q},{\rm beam}\right)$,
$\theta\left({\rm g},{\rm beam}\right)$,
$\theta\left({\rm a},{\rm beam}\right)$,
$\theta\left({\rm a},{\rm l}\right)$,
$\theta\left({\rm a},{\rm q}\right)$,
$\theta\left({\rm q},{\rm q}'\right)$,
$\theta\left({\rm l},{\rm q}\right)$,
$\theta\left({\rm l},{\rm g}\right)$,
$\theta\left({\rm l},{\rm l}'\right)$
\item energies: 
$E_{\rm l}$, $E_{\rm q}$,  $E_{\rm b}$, $E_{\rm g}$, $E_{\rm a}$
\item invariant masses: $m_{\rm ll'}$, $m_{\rm qq'}$, $m_{\rm bb'}$, 
$m_{\rm jj'}$,
\end{itemize}
where  l, l' are charged leptons, q, q' are quarks, b is a $b$-quark, 
g is a gluon, a is a photon and j is either a quark or a gluon.

The initial physical parameters: particle masses and widths, an inverse of
the fine structure constant in the Thomson limit $\alpha$, the Fermi coupling,
the strong coupling constant $\alpha_s$, number of colors and the conversion
constant can be  specified in module {\tt inprms.f}. The module contains
one more input parameter called {\tt sqscut} that is an invariant mass cut, 
which is used in mappings of the photon and gluon propagators that couple 
to massless fermions. Its value is relevant only if light fermion masses 
are chosen to be zero. If the width of the $W$ boson, Higgs boson, 
or top quark is set to zero then the program will automatically 
replace it with a value calculated in the lowest order of SM.

The helicity amplitudes are calculated with the use of the routines which have 
been collected in a directory {\tt carlolib}. The directory contains also
other routines as {\tt boost}, {\tt kinff}, {\tt kinin}, {\tt lamsq}, 
{\tt mapakk}, {\tt mapbkk}, {\tt mapskk}, which are used in the phase
space calculation, {\tt wwidth}, a routine for calculating the $W$-boson
width in the lowest order of SM, or a random number generator 
{\tt ranlux} \cite{ranlux}.

\section{Use of the program}
At present {\tt carlomat} is distributed as a single file 
{\tt carlomat.tar.gz}. After executing command \\[2mm]
{\tt tar -xzvf carlomat.tar.gz} \\[2mm]
a working directory named {\tt carlomat} with a few subdirectories is 
formed.
The part of the program responsible for code generation is stored in 
subdirectory {\tt code\_generation}.
After having specified a desired process in {\tt carlomat.f} it can be run 
with a command\\[2mm]
{\tt make test}\\[2mm]
executed in a command line in directory {\tt code\_generation}. Once ended,
a file {\tt test} is displayed on the screen that contains some details
of the process under consideration, such as the external and off-shell 
particles, the number of Feynman diagrams, etc. 
 and the generated files listed in Section 3.1
are moved to another working directory, named {\tt mc\_computation}, to which
the user should change in order to perform the MC computation.
After specifying the c.m.s. energy and the desired options 
in {\tt carlocom.f} the MC program can be run by executing a command\\[2mm]
{\tt make test}.\\[2mm]
The output of the run is stored in a file {\tt test}. 
Both files named {\tt test}, the one in a directory {\tt code\_generation}
and the other in {\tt mc\_computation} should exactly
reproduce files {\tt test0} appended to both directories.
If the MC events
generation is turned on, i.e. for {\tt imc=0,1,2,3,...}, then the generated
events are stored in a file {\tt mc\_events.x}, with {\tt x} being the c.m.s. 
energy. The data for each event are stored in {\tt mc\_events.x} in 
the following form:
\begin{center}
\begin{tabular}{rrrrr}
\multicolumn{5}{l}{\tt \# Event No.                1}\\[2mm]
{\tt p( 3)=} & 0.47046555E+02 & -0.10891034E+02& 0.42906506E+02 
& 0.15190637E+02\\
{\tt p( 4)=} & 0.12373166E+03 & 0.59317748E+02 &0.19678931E+02  
&0.10668003E+03\\
{\tt p( 5)=} & 0.25869511E+02 & 0.21415118E+02 &-0.37079632E+01 
&-0.14030830E+02\\
{\tt p( 6)=} & 0.17716225E+03 &-0.40997709E+02 & 0.81425324E+01 
&-0.17216082E+03\\
{\tt p( 7)=} & 0.65263801E+02 &-0.52555950E+02 &-0.20628053E+02 
& 0.32737122E+02\\
{\tt p( 8)=} & 0.60926224E+02 & 0.23711829E+02 &-0.46391953E+02 
& 0.31583866E+02
\end{tabular}
\end{center}
where {\tt p(3),...,p(8)} are four momenta of the final state particles in the
centre of mass system in exactly the same order as they appear in
{\tt process}. The first component is the energy and the other
3 components are the $x,y$ and $z$ components of the four momentum.
The number of accepted events substantially increases when the program 
is run the second time with
exactly the same choice of options and parameters. 
In each run, the maximum values of the cross section are stored in a file
named {\tt csmax\_*} in directory {\tt mc\_computation} and they are 
automatically read from the disc in the next run of the program.
The efficiency of events acceptance is not high, but it can be improved 
by a proper selection
of the differential cross section and an appropriate choice of the number
of bins.
\section{Outlook}
The following improvements of the program are envisaged in the near future.
\begin{itemize}
\item Dedicated treatment of soft and collinear external photons,
      as well as $t$-channel photon/gluon exchange.
\item Interfaces to parton density functions, or the initial state radiation 
      within the structure function approach.
\item More efficient calculation of colour coefficients.
\end{itemize}
Moreover, interfaces to parton shower and hadronization programs should 
be worked on. Extensions of SM can be implemented and the corresponding 
lowest order cross sections can be calculated in a fully automatic way.
Leading SM radiative corrections can be implemented, if
the necessary subroutines are provided.


%

\end{document}